\documentclass[preprint,12pt]{aastex}
\usepackage{amsmath}

\shorttitle{Diagnosing the Outflow from SGR 1806--20 Giant Flare}

\begin{document}
\title{Diagnosing the Outflow from the SGR 1806--20 Giant Flare with Radio Observations}

\author{J. Granot\altaffilmark{1}, E. Ramirez-Ruiz\altaffilmark{2},
G.~B.~Taylor\altaffilmark{1,3}, D.~Eichler\altaffilmark{4},
Y.~E.~Lyubarsky\altaffilmark{4}, R.~A.~M.~J.~Wijers\altaffilmark{5},
B.~M.~Gaensler\altaffilmark{6}, J.~D.~Gelfand\altaffilmark{6},
C.~Kouveliotou\altaffilmark{7}}

\altaffiltext{1}{KIPAC, Stanford University, P.O. Box 20450, MS 29,
Stanford, CA 94309; granot@slac.stanford.edu}
\altaffiltext{2}{Institute for Advanced Study, Einstein Drive,
Princeton, NJ 08540; Chandra Fellow} \altaffiltext{3}{National Radio
Astronomy Observatory, P.O. Box O, Socorro, NM 87801 }
\altaffiltext{4}{Department of Physics, Ben Gurion University,
P.O. Box 653, Be'er Sheva 84105, Israel} \altaffiltext{5}{Astronomical
Institute ``Anton Pannekoek'', University of Amsterdam, Kruislaan 403,
1098 SJ, Amsterdam, The Netherlands}
\altaffiltext{6}{Harvard-Smithsonian Center for Astrophysics, 60
Garden Street, Cambridge, MA 02138} \altaffiltext{7}{NASA/Marshall
Space Flight Center, XD-12, NSSTC, 320 Sparkman Dr., Huntsville, AL
35805, USA}

\begin{abstract}

On 2004 Dec. 27, the soft gamma repeater (SGR) $1806-20$ emitted the
brightest giant flare (GF) ever detected from an SGR. This burst of
energy, which resulted in an (isotropic) energy release $\sim 100$
times greater than the only two other known SGR GFs, was followed by a
very bright, fading radio afterglow.  Extensive follow-up radio
observations provided a wealth of information with unprecedented
astrometric precision, revealing the temporal evolution of the source
size, along with densely sampled light curves and spectra. Here we
expand on our previous work on this source, by explaining these
observations within one self-consistent dynamical model.  In this
scenario, the early radio emission is due to the outflow ejected
during the GF energizing a thin shell surrounding a pre-existing
cavity, where the observed steep temporal decay of the radio emission
seen beginning on day 9 is attributed to the adiabatic cooling of the
shocked shell. The shocked ejecta and external shell move outward
together, driving a forward shock into the ambient medium, and are
eventually decelerated by a reverse shock.  As we show in Gelfand et
al. (2005), the radio emission from the shocked external medium
naturally peaks when significant deceleration occurs, and then decays
relatively slowly.  The dynamical modeling of the collision between
the ejecta and the external shell together with the observed evolution
of the source size (which is nicely reproduced in our model) suggest
that most of the energy in the outflow was in mildly relativistic
material, with an initial expansion velocity $v/c\lesssim 0.7d_{15}$,
for a distance of $15d_{15}\;$kpc to SGR~$1806-20$.  An initially highly
relativistic outflow would not have produced a long coasting phase at
a mildly relativistic expansion velocity, as was observed.

\end{abstract}


\section{Introduction}

Soft gamma repeaters (SGRs) are believed to be ``magnetars'' -- a
small class of slowly spinning neutron stars with extremely high
surface magnetic fields, $B\ga 10^{15}\;$G \citep{DT92,K98}. They have
been discovered through their transient X-ray outbursts, during which
they emit hundreds of short ($\sim 0.2\;$s), soft ($kT\sim25\;$keV)
bursts; very rarely (only twice so far), SGRs emit giant flares (GF),
extreme events with luminosities upwards of $10^{44}\;{\rm
erg\;s^{-1}}$. SGR~$1806-20$ lies in the Galactic plane, at a distance
of about $d=15\,d_{15}\;$kpc \citep{CE04,Cameron05,M-GG05}.  The third
GF yet recorded occurred on 27 Dec. 2004, when SGR~$1806-20$ emitted a
burst so extreme that it was the brightest extra-solar transient event
ever recorded \citep{Palmer05,Hurley05}.  The flare was also unique in
creating a very bright radio afterglow \citep{Gaensler05,CK05}, which
was monitored for months, providing an amazing wealth of data at
several radio frequencies, including the temporal evolution of the
source size and shape, polarization and flux.

The data from the radio source that appeared in the aftermath of the
GF provide a rare opportunity for a detailed study of a mildly
relativistic blast wave, which might help bridge the gap between the
relativistic outflows in cosmological gamma-ray bursts and supernova
remnants. \citet{Gelfand05} present a rebrightening episode in the
radio light curve, which is well fit by a semi-analytic spherical
model where the radio emission resulted from a blast wave driven by
$\ga 10^{24.5}\;$g of baryonic material driven off the neutron star,
and the expanding radio nebula has now entered its Sedov-Taylor phase
of evolution. An accompanying study of the evolution of the size of
the nebula confirms that it is indeed decelerating
\citep{Taylor05}. Furthermore, the motion of the flux centroid implies
a predominantly one-sided mildly collimated outflow, i.e. a wide
one-sided jet \citep{Taylor05}.

In this paper, we expand on the framework laid out by
\citet{Gaensler05}, \citet{Gelfand05} and \citet{Taylor05}, and
present a full dynamical model for the interaction of the outflow that
was ejected during the 2004 Dec.  27 GF with its surroundings, and in
particular with an external shell. Our model, which is described in \S
\ref{dynamics}, explains the large and diverse data sets for this
event and constrains the initial velocity of the ejecta from the
flare. Both a relativistic (\S \ref{rel}) and a Newtonian (\S
\ref{Newt}) outflow are considered. We find that only a Newtonian
outflow with an initial expansion velocity $v\approx 0.7\,d_{15}c$
fits the observations well.  In \S \ref{emission} we derive the
synchrotron emission implied by our dynamical model and show that it
also agrees nicely with the radio observations. Our conclusions are
discussed in \S \ref{diss}.

\section{The Underlying Dynamics}
\label{dynamics}

The radio light curve initially exhibited a relatively moderate decay,
$\sim t^{-1.5}$, followed by an achromatic steepening at $t_b\approx
9\;$days after the GF, to $\sim t^{-2.7}$ \citep{Gaensler05}. This was
followed by a rebrightening starting at $\sim 25\;$days and peaking at
$t_p\approx 33\;$days \citep{Gelfand05}; the decay rate slowed
significantly thereafter ($\sim t^{-1.1}$; Gelfand et al. in
preparation).

The apparent expansion velocity, $v_{\rm ap}=\beta_{\rm ap}c$, of the
nebula was initially fairly constant,\footnote{We adopt the value that
was derived by Taylor et al. (2005) for the average expansion velocity
during the first $82\;$days, which is slightly higher than the value
reported initially by \citet{Gaensler05}.} $\sim 0.4\,d_{15}c$, and
decreased at $t_{\rm dec}\sim t_p$ \citep{Taylor05}.  This value of
$v_{\rm ap}$ is for the geometrical mean of the semi-major axis and
semi-minor axis, and assumes a double sided outflow. Under the latter
assumption, $v_{\rm ap}$ along the semi-major axis and semi-minor axis
in $\sim 0.5\,d_{15}c$ and $\sim 0.25\,d_{15}c$,
respectively. However, as the motion of the flux centroid implies that
the SGR is located close to the edge of the image along the semi-major
axis \citep{Taylor05}, this implies $v_{\rm ap}\sim 1.0\,d_{15}c$ for
the leading edge of the one-sided outflow.

The true velocity (in units of $c$) of the emitting material near the
edge of the image is $\beta=(1-\Gamma^{-2})^{1/2}=\beta_{\rm
ap}/(\sin\theta+\beta_{\rm ap}\cos\theta)$ where $\theta$ is the angle
between its direction of motion and the line of sight. The minimal
true velocity corresponding to a given observed value of $\beta_{\rm
ap}$ is $\beta_{\rm min}=\beta_{\rm ap}(1+\beta_{\rm ap}^2)^{-1/2}$ or
$\Gamma_{\min}\beta_{\rm min}=\beta_{\rm ap}$ and is obtained for an
angle $\theta_{\rm min}$ that satisfies $\cos\theta_{\rm
min}=\beta_{\rm min}$. The same angle, $\cos\theta=\beta$, is where
$\beta_{\rm ap}$ is maximal for a fixed $\beta$ (and therefore
$\beta_{\rm ap}=\Gamma\beta$ at the outer edge of the image of a
spherical source expanding at a velocity $\beta$). Therefore, we
expect $\Gamma\beta\sim\beta_{\rm ap}$, and in our case
$\Gamma\beta\sim 1.0\,d_{15}$, so that $\beta\sim
d_{15}(1+d_{15}^2)^{-1/2}$ or $\beta\sim 0.7$ for $d_{15}\approx 1$.

Since the axis ratio of the radio image at the times relevant for our
modeling is at most $\sim 2:1$, and for the sake of simplicity, we
adopt a spherical model for most of our analysis, and later introduce
the corrections for a mildly collimated one sided outflow, as is
implied by the observations.

As the source was already fading by the time of the first observation
($t_{\rm I}\approx 7\;$days), the radio emission must have turned on
at an earlier time and at a smaller radius. The radio spectrum and
linear polarization observations suggest that we are seeing
synchrotron emission.  If the relativistic electrons that are emitting
this radiation were accelerated at a much smaller radius ($\ll
10^{16}\;$cm) then most of their energy would have been lost via
adiabatic cooling by $t_{\rm I}$, thus dramatically increasing the GF
energy requirements. Moreover, the achromatic light curve steepening
at $t_b$ \citep{Gaensler05}, strongly suggests a hydrodynamic
transition at that time. A simple explanation for this overall
behavior arises if the outflow from the GF initially coasted with
negligible interaction with the ambient medium, until at $t_{\rm col}=
5\,t_{\rm col,5}\;$days it collided with a thin external shell, which
caused it to decelerate by a reverse shock, while the external shell
was accelerated by a forward shock.\footnote{Such a shell surrounding
a pre-existing cavity is thought to be formed behind the bow shock due
to the supersonic motion of SGR~$1806-20$ through the ISM and its
quiescent wind \citep{Gaensler05}.  Alternatively, it could also arise
from an earlier and initially faster mass ejection from the
SGR~$1806-20$ GF, which was decelerated by the external medium to a
velocity slightly below that of the coasting second shell, thus
causing the two shells to collide \citep[the ``refreshed shock''
scenario, e.g.][]{GNP03}.}  After this collision the two shells move
together at a somewhat reduced speed. Thus the emission up to $t_b$ is
dominated by the {\it recently} shocked electrons in these two
shells. The radiation then arrives at the observer at a time
$t\lesssim 2\,t_{\rm col}$ due to light travel effects and the
finite time it takes for the shock to cross the shells. At $t>t_b$ the
emission is dominated by the adiabatically cooling electrons in the
two shells. As shown in \S \ref{emission} \citep[see
also][]{Gaensler05} this naturally produces the unusually steep decay
in the light curve.

As the merged shell expands outwards it drives a shock into the
ambient medium. An increasing amount of external mass is swept up,
until the emission from this shocked fluid starts dominating the light
curve at $t\gtrsim 25\;$days. This naturally produces a rebrightening
in the light curve which peaks at $t_p\approx 33\;$days
\citep{Gelfand05}; as expected, a decrease in the expansion velocity
was observed at about the same time, $t_{\rm dec}\sim t_p$ (Taylor et
al. 2005). At $t>t_{\rm dec}$ the hydrodynamics gradually approach the
self-similar Sedov-Taylor solution, which predicts a slower flux decay
rate, in agreement with observations \citep{Gelfand05}. An outline of
our basic picture is shown in panel {\bf a} of Fig. \ref{model}. Below
we reproduce the main observed features using a simple analytic model
for the interaction between the outflow and its surroundings, and
present a numerical simulation that broadly agrees with the analytic
model and nicely reproduces the observed evolution of the source size.

\subsection{Relativistic Outflow}
\label{rel}

A simple model for the collision between the cold ejecta shell of
initial Lorentz factor $\Gamma_{\rm ej}=(1-\beta_{\rm ej}^2)^{-1/2}$
and mass $M_{\rm ej}$, and an external thin shell of mass $M_{\rm
ext}$ at rest at a radius $R_{\rm ext}$, is a plastic collision where
the two shells are shocked (the two shocked fluids separated by a
contact discontinuity) and subsequently move together at
$\Gamma_f=(1-\beta_f^2)^{-1/2}$.  Both $\Gamma_{\rm ej}$ and
$\Gamma_f$ are measured in the rest frame of the unperturbed external
medium.  Energy and momentum conservation in the rest frame of the
merged shell require $E_f/c^2=M_f=\Gamma_r M_{\rm ej} + \Gamma_f
M_{\rm ext}$ and $\Gamma_r\beta_r M_{\rm ej}=\Gamma_f\beta_f M_{\rm
ext}$, respectively, where $\Gamma_r =
(1-\beta_r^2)^{-1/2}=\Gamma_{\rm ej}\Gamma_f(1-\beta_{\rm ej}\beta_f)$
is the initial Lorentz factor of the ejecta in the rest frame of the
merged shell.  The resulting internal energy is $E_{\rm
int}/c^2=(\Gamma_r-1)M_{\rm ej}+(\Gamma_f-1)M_{\rm ext}$ and the final
velocity is
\begin{equation}\label{beta_f}
\frac{\beta_f}{\beta_{\rm ej}}=\left(1+\frac{M_{\rm
ext}}{\Gamma_{\rm ej}M_{\rm ej}}\right)^{-1}\ .
\end{equation}
This shows that an external mass of $M_{\rm ext}\sim\Gamma_{\rm
ej}M_{\rm ej}$ is required in order to significantly reduce the
initial velocity.

For an {\it initially} relativistic outflow $\Gamma_{\rm ej}\approx
E/M_{\rm ej}c^2\gg 1$ and $\beta_{\rm ej}\approx 1$ so that
$\beta_f\approx(1+M_{\rm ext}c^2/E)^{-1}$, which for $M_{\rm
ext}c^2\gg E$ (and correspondingly $\beta_f\ll 1$) gives $M_{\rm ext}
v_f^2\approx\beta_f E$.  Therefore, in this limit, the kinetic energy
of the merged shell carries only a small fraction ($\sim\beta_f$) of
the total energy, while most of the energy is in the internal energy
of the shocked ejecta ($E_{\rm int}\approx E \approx \Gamma_{\rm ej}
M_{\rm ej} c^2$).  The relativistically hot shocked ejecta can then
convert most of its internal energy back into kinetic energy through
$PdV$ work as the merged shell keeps expanding. This might initially
(soon after the collision) accelerate the shell, and later cause it to
decelerate more slowly with time (and radius), thus increasing the
radius, $R_{\rm dec}$, where it decelerates significantly, compared to
its value for a cold shell with the same (post collision) mass and
velocity, $R_{\rm dec}\sim 2^{1/3}R_{\rm ext}$.

Nevertheless, even if all the original energy is back in the form of
kinetic energy at $R_{\rm dec}$, then still $E/c^2\approx M_{\rm
dec}\beta_{\rm dec}^2\approx\beta_f M_{\rm ext}$ where $M_{\rm
dec}=M(R_{\rm dec})\approx(4\pi/3)\rho_{\rm ext}R_{\rm dec}^3$ is the
total mass (in the shells and swept-up external medium) at $R_{\rm
dec}$, and $\beta_{\rm dec}=\beta(R_{\rm dec})\approx
0.4\,d_{15}\gtrsim\beta_f$. Finally, $M_{\rm
ext}\approx(4\pi/3)\rho_{\rm ext}R_{\rm ext}^3$ for most reasonable
scenarios that produce an external shell, such as a bow shock
\citep{Wilkin96}.\footnote{This is an important assumption.  If
somehow the mass of the shell would be larger by some factor
$f=3M_{\rm ext}/4\pi\rho_{\rm ext}R_{\rm ext}^3$, then $R_{\rm
dec}/R_{\rm ext}$ would increase by a factor of $f^{1/3}$ so that
$f\sim 10^2$ would be required in order to explain the observed
evolution of the source size.  Therefore, an external shell with
$f\sim 10^2$, or alternatively a sharp density drop by a factor of
$\sim 10^2$ around $R_{\rm ext}$ that lasts for at least an order of
magnitude in radius, would in principle be consistent with the
observations. In practice, however, such external density profiles
seem highly contrived and therefore not very likely.} This gives
\begin{equation}\label{mass_ratio}
\left(\frac{R_{\rm dec}}{R_{\rm ext}}\right)^3\approx \frac{M_{\rm
dec}}{M_{\rm ext}}\approx \frac{\beta_f}{\beta_{\rm
dec}^2}\lesssim\beta_{\rm dec}^{-1} \sim 2.5\,d_{15}^{-1}\ .
\vspace{0.25cm}
\end{equation}

We now proceed to compare the radio observations with the above
calculations.  The angular diameter of the source at the time of the
first observation, $t_{\rm I}\approx 7\;$days, and at the epoch of
deceleration, $t_{\rm dec}\sim t_p\approx 33\;$days, was about
$80\;$mas and $300\;$mas, respectively.\footnote{At both epochs the
image is somewhat elongated and the quoted value is along the
semi-major axis \citep{Gaensler05,Taylor05}. The ratio of the angular
size at these two epochs, however, is rather robust and a comparable
ratio is obtained along the semi-minor axis. This ratio is also
applicable for a one sided outflow, as suggested in \citet{Taylor05}.}
The corresponding radii are $R_{\rm I}=9.0\times 10^{15}d_{15}\;$cm
and $R_{\rm dec}\approx 3.4\times 10^{16}d_{15}\;$cm. The requirement
that $R_{\rm ext}<R_{\rm I}$ gives $R_{\rm dec}/R_{\rm ext}\gtrsim
3.75$ and therefore $(R_{\rm dec}/R_{\rm ext})^3\approx 50(R_{\rm
I}/R_{\rm ext})^3\gtrsim 50$, which contradicts Eq.  \ref{mass_ratio}
for any reasonable value of $d_{15}$. Thus, an ultra-relativistic
outflow ($\Gamma_{\rm ej}\gg 1$) fails to reproduce the observations,
since $R_{\rm dec}$ would not be much larger than $R_{\rm ext}$;
specifically we expect $R_{\rm dec}/R_{\rm ext}\lesssim
1.4d_{15}^{-1/3}$ (see Eq. \ref{mass_ratio}).

\subsection{Newtonian Outflow}
\label{Newt}

For a Newtonian outflow ($\beta_{\rm ej}\ll 1$), Eq. \ref{beta_f}
reduces to $\beta_f/\beta_{\rm ej}\approx M_{\rm ej}/M_f$, where
$M_f\approx M_{\rm ej}+M_{\rm ext}$. Since $M(R_{\rm ext}<r<R_{\rm
dec})\sim M_f$ and $M_{\rm ext}\approx(4\pi/3)\rho_{\rm ext}R_{\rm
ext}^3$, then $M_{\rm ext}>M_{\rm ej}$ would imply $R_{\rm dec}/R_{\rm
ext}\sim 2^{1/3}\approx 1.26$, in contrast with observations.

Therefore, we must have $M_{\rm ej}\gg M_{\rm ext}$, which results in
$\beta_f\approx\beta_{\rm ej}$, $M_f\approx M_{\rm ej}\sim M_{\rm
dec}\approx(4\pi/3)\rho_{\rm ext}R_{\rm dec}^3$, and $M_{\rm
ej}/M_{\rm ext}\approx(R_{\rm dec}/R_{\rm ext})^3\gtrsim 50$, or
$R_{\rm ext}\approx(t_{\rm col}/t_{\rm I})R_{\rm I}\approx 6.4\times
10^{15}t_{\rm col,5}d_{15}\;$cm and $M_{\rm ej}/M_{\rm
ext}\approx(R_{\rm dec}/R_{\rm ext})^3\approx 140t_{\rm col,5}^{-3}$.
At $t_{\rm dec}$ we directly measure the source size, $R_{\rm dec}$,
and expansion velocity, $\beta_{\rm ej}\approx\beta_f\approx R_{\rm
dec}/ct_{\rm dec}\approx 0.4d_{15}$. Therefore, since the shocked
external medium has comparable internal and kinetic energies, the
energy in the outflow is given by $E\approx(4\pi/3)\rho_{\rm
ext}R_{\rm dec}^3v_{\rm ej}^2\approx 3.8\times 10^{46}n_0
d_{15}^5\;$erg, and depends only on the unknown external density,
$n_{\rm ext}=\rho_{\rm ext}/m_p=n_0\;{\rm cm^{-3}}$. Here $v_{\rm
ej}=\beta_{\rm ej}c$ and $E_{46}=E/(10^{46}\;$erg). Thus $n_0\approx
0.26d_{15}^{-5}E_{46}$ and $M_{\rm ej}\approx E/v_{\rm ej}^2\approx
2.7\times 10^{26}n_0 d_{15}^3\;$g. These results for $E$ and $M_{\rm
ej}$ are similar to those derived by \citet{Gelfand05}.

A simple generalization for a wide one-sided jet is as follows.  The
volume of the shocked external fluid and therefore its mass for a
given external density does not change. The kinetic energy per unit
rest energy, $\Gamma-1$, grows by a factor of $\sim 4-5$ at the head
of the jet (where $\Gamma\beta\approx 1$, see beginning of \S
\ref{dynamics}) and decreases near the SGR. On average it increases by
a factor of $\sim 2-3$, and the estimate for $E/n_{\rm ext}$ increases
by the same factor while $M_{\rm ej}/n_{\rm ext}$ does not change.

Finally, it is important to keep in mind that the outflow might
consist of more than one component. The simplest example is a
relativistic shell (with $\Gamma\gg 1$) followed by a Newtonian shell
(with $\Gamma\beta\lesssim 1$) that was ejected slightly later during
the GF. The relativistic shell is shocked and decelerated to a
Newtonian velocity as it collides with the external shell, at $t_{\rm
col,1}$, while the Newtonian shell catches up and collides with the
slower merged relativistic + external shell at $t_{\rm col,2}>t_{\rm
col,1}$. As long as the velocity after the first collision is
sufficiently smaller than that of the Newtonian shell, the subsequent
dynamics would not be very different than for the Newtonian outflow
case discussed above. An important difference, however, is that the
emission would light up at $\sim t_{\rm col,1}/2\Gamma^2\ll t_{\rm
col,2}$, i.e. much earlier than without the relativistic component. (A
similar result is obtained also if there is a continuous external
medium instead of a shell surrounding a cavity.)  Rapid follow-up
observations of future GFs could test this hypothesis directly, and
teach us more about the properties of the outflow. In the present
case, the later collision with the Newtonian shell might explain the
change in the degree of linear polarization (from decreasing to
increasing with time) and its position angle, at $t\approx 10\;$days
\citep{Gaensler05}.

We have tested the colliding shell scenario with the aid of numerical
simulations, which model the dynamics much more accurately than the
simple analytic model used above. Our basic picture is confirmed by
these calculations, and the observed evolution of the source size is
nicely reproduced (see Fig. \ref{hydro}). The simple analytic
expression for the source size is $R(t>t_{\rm col})\approx R_{\rm
dec}\min[(t/t_{\rm dec}),(t/t_{\rm dec})^{2/5}]$. The two asymptotic
power laws correspond to the coasting phase and the Sedov-Taylor
regime. A semi-analytic model which gives a smooth transition between
these two asymptotic power laws is presented in Fig. 2 of
\citet{Gelfand05} and fitted to the source size as a function of time
in Fig. 2 of \citet{Taylor05}. Around the time of the collision,
$t\sim t_{\rm col}$, there is a flat part in $R(t)$ due to the shocks
that are crossing both shells. This flat part is nicely reproduced by
the numerical simulation, and its exact shape depends on the details
of the collision (and thus does not have a simple and robust analytic
description). In a future work (Ramirez-Ruiz et al., in preparation)
we investigate the dynamics in more detail, including the implications
of aspherical outflows, that are relevant given the elongated nature
of the radio image \citep{Gaensler05} and the motion of its flux
centroid \citep{Taylor05}.

\section{Explaining the Observed Radio Emission}
\label{emission}

Once the reverse shock crosses the ejecta and the forward shock
crosses the external shell, the supply of newly accelerated electrons
will be exhausted. As the merged shells expand, the existing
relativistic electrons cool adiabatically and the magnetic field
decreases, thus nicely reproducing the sharp decay that was observed
in the radio flux between $9$ and $\sim 25\;$days, $\sim t^{-2.7}$
\citet{Gaensler05}.

The emission from the forward shock is dominated by the newly shocked
electrons which are accelerated to relativistic energies with a power
law distribution, $dN/d\gamma_e\propto\gamma_e^{-p}$ for
$\gamma_e>\gamma_m$. At $t<t_{\rm dec}$ the relative velocity of the
shocked downstream fluid and the upstream fluid is roughly constant
and equal to $v_{\rm rel}=\beta_{\rm rel}c\approx 0.3d_{15}c$ since
$v_{\rm rel}/v_{\rm ej}\approx 3/4$. The average Lorentz factor of the
relativistic electrons is given by $\langle\gamma_e\rangle=\epsilon_e
e'/\xi_en'm_ec^2$ where $e'$ and $n'$ are the proper internal energy
density and number density of the shocked fluid, $\epsilon_e$ is the
fraction of the post shock internal energy density in relativistic
electrons, and $\xi_e$ is the fraction of electrons that are
accelerated to relativistic energies. The energy per proton is
$e'/n'=(\Gamma_{\rm rel}-1)m_pc^2\approx(\beta_{\rm rel}^2/2)m_pc^2$,
where the second expression is valid in the limit of a Newtonian blast
wave. For $p>2$ we have $\gamma_m=\langle\gamma_e\rangle(p-2)/(p-1)$
and therefore
\begin{equation}\label{gamma_m}
\gamma_m=\frac{\epsilon_e}{\xi_e}\left(\frac{p-2}{p-1}\right)
\frac{m_p}{m_e}\frac{\beta_{\rm rel}^2}{2}=
2\,g\,\xi_e^{-1}\epsilon_{e,-1}\left(\frac{\beta_{\rm rel}}{0.26}\right)^2\ ,
\end{equation}
where $g=3(p-2)/(p-1)$ ($=1$ for $p=2.5$), and
$\epsilon_{e,-1}=\epsilon_e/0.1$. Since $\gamma_m$ is the lowest
Lorentz factor of the {\it relativistic} electrons, by definition
$\gamma_m\gtrsim 2$.  \citet{Gelfand05} calculate the light curve
under the assumption that $\epsilon_e/\xi_e={\rm const}$ and
$\gamma_m>2$ \citep[see also][]{FWK00}, which is valid for
$\epsilon_e>0.1$ or $\xi_e\ll 1$ until there is significant
deceleration.  Once $\gamma_m$ decreases to $\sim 2$, the subsequent
behavior of $\epsilon_e$ and $\xi_e$ depends on poorly understood
shock acceleration of non-relativistic electrons.  Here it is assumed
that $\epsilon_e={\rm const}$.  Eq.  \ref{gamma_m} shows that for
$\epsilon_{e,-1}\lesssim 1$ (and it is difficult for $\epsilon_e$ to
be much higher than $0.1$) we have $\gamma_m\sim 2$ which is constant
all along, while $\xi_e\sim(v_{\rm rel}/v_{\rm ej})^2\propto\beta_{\rm
rel}^2$ decreases with time at $t\gtrsim t_{\rm dec}$. This results in
a more moderate temporal decay of the flux at $t>t_{\rm dec}$ (see
Eq. \ref{F_nu}), in better agreement with observations.

At $t\gg t_{\rm dec}$ the shock dynamics approach the Sedov-Taylor
self-similar solution, where $R\approx(Et^2/\rho_{\rm
ext})^{1/5}$. Therefore, $v_{\rm rel}/v_{\rm ej}\approx(t_{\rm
dec}/t)R/R_{\rm dec}\approx\min[1,(t/t_{\rm dec})^{-3/5}]$ and $v_{\rm
sh,0}/v_{\rm ej}\approx v_{\rm sh,0}/v_f=4/3$ where $t_{\rm
dec}=R_{\rm dec}/v_{\rm sh,0}=(3E/2\pi\rho_{\rm ext}v_{\rm
sh,0}^5)^{1/3}$. Here $v_{\rm sh,0}$ is the initial velocity of the
shock front for the blast wave propagating into the external medium.
The post shock magnetic field is $B=(8\pi\epsilon_Be_{\rm int})^{1/2}$
where $e_{\rm int}=2\rho_{\rm ext}v_{\rm rel}^2$ where
$\epsilon_B=0.1\epsilon_{B,-1}$ is the fraction of the post shock
internal energy in the magnetic field.  The number of synchrotron
emitting electrons is $N_e=\xi_e M/m_p$ where $M=f_b(4\pi/3)\rho_{\rm
ext}R^3$ and $f_b$ is the beaming factor (i.e. the fraction of the
total solid angle occupied by the outflow). Finally,
$F_{\rm\nu,max}=N_eP_{\rm\nu,max}/4\pi d^2$, where
$P_{\rm\nu,max}\approx P_{\rm syn}/\nu_{\rm syn}$, $P_{\rm
syn}(\gamma_e)=(4/3)\sigma_Tc(B^2/8\pi)\gamma_e^2$, and $\nu_{\rm
syn}(\gamma_e)=eB\gamma_e^2/2\pi m_e c$. The observed spectral slope
in the radio suggest that we are in the spectral power law segment
$\nu_m<\nu<\nu_c$, where $\nu_m=\nu_{\rm syn}(\gamma_m)$ and $\nu_c$
is the cooling break frequency. Thus we find
\begin{equation}\label{F_nu}
F_\nu = 4.2\,f_b\,g\, n_0^{3(p+1)/20}E_{46}^{(11+p)/10}\epsilon_{e,-1}
\left(\frac{\epsilon_B}{0.002}\right)^{(p+1)/4} 
d_{15}^{-2} \left(\frac{\nu}{8.5\,{\rm GHz}}\right)^{(1-p)/2}
\left(\frac{t}{33\,{\rm days}}\right)^{-3(p+1)/10}\;{\rm mJy}\ \quad
\end{equation}
at $t>t_{\rm dec}$, while $F_\nu(t<t_{\rm dec})\approx (t/t_{\rm
dec})^3F_\nu(t_{\rm dec})$.

The parameter values in Eq. \ref{F_nu} were chosen to match the
observed flux at the peak of the rebrightening, $t_{\rm dec}\approx
t_p\approx 33\;$days. This demonstrates that an energy of $\sim f_b
10^{46}\;$erg, comparable to that in the GF (if it were emitted into a
similar solid angle as the outflow), can be accommodated for
reasonable values of the micro-physical parameters and the external
density.  Taking into account the relation $n_0\approx
0.26d_{15}^{-5}E_{46}$ derived in \S \ref{Newt}, we find that an
equipartition limit $\epsilon_e,\epsilon_B\lesssim 0.3$ ($0.5$) gives
$E_{44}\gtrsim 7.5\,d_{15}^{2.5}$ ($4.0\,d_{15}^{2.5}$), where
$E_{44}=E/(10^{44}\;{\rm erg})$, consistent with the conclusions of
\citet{Gelfand05}.
For a wide one sided jet of half-opening angle $\theta_0\approx
0.5\;$rad, $f_b=(1-\cos\theta_0)/2\approx 0.06$, while $E/n_{\rm ext}$
grows by a factor of $\sim 2^5$ where $E$ is the isotropic equivalent
energy. Altogether, we obtain for the true energy $E_{44}\gtrsim
5.7\,d_{15}^{2.5}$ ($3.4\,d_{15}^{2.5}$), as well as $n_0\gtrsim
7.4\times 10^{-3}d_{15}^{-2.5}$ ($4.4\times 10^{-3}d_{15}^{-2.5}$) and
$M_{\rm ej,24}\gtrsim 9.9\,d_{15}^{0.5}$ ($5.9\,d_{15}^{0.5}$), where
$M_{\rm ej,24}=M_{\rm ej}/(10^{24}\;{\rm g})$.

Finally, we estimate the expected flux at the end of the collision
between the ejecta and the external shell, at $\sim t_{\rm col}$. The
external shell is accelerated to $\beta_f\approx\beta_{\rm ej}$ while
the ejecta are only slightly decelerated, so that the shock going into
the external shell is stronger and likely to dominate the
emission. The volume of the shell, $4\pi\eta R_{\rm ext}^3$ where
$\eta=\Delta R/R_{\rm ext}=0.1\eta_{-1}$, is reduced by a factor of
$4$ due to shock compression, and its internal energy is a fraction
$M_{\rm ext}/M_f\approx M_{\rm ext}/M_{\rm ej}\approx 0.007t_{\rm
col,5}^3$ of the total energy $E$. Under similar assumptions as above,
\begin{equation}\label{F_nu2}
F_\nu(t_{\rm col}) \approx 80\,f_b\,g^{p-1}\,\eta_{-1}^{-(p+1)/4}
E_{46}^{(5+p)/4}\epsilon_{e,-1}^{p-1}\epsilon_{B,-1}^{(p+1)/4}
d_{15}^{(5p-27)/4} \left(\frac{\nu}{8.5\,{\rm
GHz}}\right)^{(1-p)/2}t_{\rm col,5}^{3} \;{\rm mJy}\ ,
\end{equation}
in rough agreement with the extrapolation to $t_{\rm col}\sim 5\;$days
of the observed flux, $F_{\rm\nu=8.5\,{\rm GHz}} = 53\;$mJy, at the
first epoch, $t_{\rm I}=6.9\;$days \citep{CK05,Gaensler05}.

For the parameter values used in Eq. \ref{F_nu2} we obtain $\nu_m\sim
1\;$MHz, $\nu_{\rm sa}\sim 50\;$MHz and $\nu_c\sim 10^{17}\;$Hz at
$t_{\rm col}\sim 5\;$days, where $\nu_{\rm sa}$ is the self absorption
frequency, so that the radio frequencies are well within the assumed
power law segment of the spectrum. The low value we obtain for
$\nu_{\rm sa}$ is consistent with the lack of a change in the spectral
slope down to $240\;$MHz \citep{Cameron05}. Soon after the shock
finishes crossing the shell, the electron power law energy
distribution extends up to $\gamma_{\rm max}\sim\gamma_c(t_{\rm
col})\sim 10^6$. Thereafter adiabatic cooling takes over and
$\gamma_{\rm max}\propto t^{-2/3}$, while $B\propto t^{-1}$ so that
$\nu_{\rm syn}(\gamma_{\rm max})\sim\nu_c(t_{\rm col})(t/t_{\rm
col})^{-7/3}$. The emission from the shocked external medium starts to
dominate at $t\approx 25\;$days, i.e. $t/t_{\rm col}\sim 5$, and hence
at that time $\nu_{\rm syn}(\gamma_{\rm max})\gtrsim 10^{15}\;$Hz is
well above the radio.

\section{Discussion}
\label{diss}

We have described a dynamical model for the interaction with the
surrounding medium of the outflow during the 2004 Dec. 27 giant flare
(GF) from SGR~$1806-20$.  This model nicely accounts for the observed
radio light curves and spectrum as well as for the evolution of the
source size with time. Using a simple analytic model, we have shown
that the bulk of outflow from the GF could not have been highly
relativistic, and was instead only mildly relativistic, with an
average velocity of $v\lesssim 0.7d_{15}c$, similar to the observed
roughly constant expansion velocity over the first month or so, taking
into account the one-sided wide jet that is suggested by the radio
data \citep{Taylor05}.\footnote{The local expansion velocity is
highest along the jet axis. The observed axis ratio of $\sim 2:1$ for
the radio image \citep{Gaensler05,Taylor05} sets a lower limit on the
true axis ratio of the emitting region.} 

A major ingredient in our dynamical model is an external shell at a
distance of $\sim 10^{16}\;$cm from the SGR. Such an external shell
might naturally be formed by the bow shock due to the SGR's quiescent
wind and its supersonic motion relative to the external medium
\citep{Gaensler05}. A bow shock origin of the external shell has
interesting implications. The bow shock stand-off radius is $R_{\rm
bs}=9.0\times 10^{15}L_{34.5}^{1/2}n_0^{-1/2}v_{250}^{-1}\;$cm, where
$v_*=250v_{250}\;{\rm km\;s^{-1}}$ is the velocity of SGR~$1806-20$
relative to the external medium, and $L=10^{34.5}L_{34.5}\;{\rm
erg\;s^{-1}}$ is its spin-down luminosity.\footnote{Before 1999
$L\approx 8\times 10^{33}\;{\rm erg\;s^{-1}}$ while by 2001 and until
before the Dec. 27th GF it leveled off at $L\approx 4.5\times
10^{34}\;{\rm erg\;s^{-1}}$ (Woods et al. in preparation).  The
dynamical time scale for the bow shock is $t_{\rm bs}\sim R_{\rm
bs}/v_*=11.4L_{34.5}n_0^{-1/2}v_{250}^{-2}\;$yr. In our scenario,
$R_{\rm bs}\sim (1-2)R_{\rm ext}$ so that $t_{\rm bs}\sim 12t_{\rm
col,5}d_{15}v_{250}^{-1}\;$yr. Since the spin down rate of
SGR~$1806-20$ increased by a factor of $\sim 5$ several years before
the GF, the steady state assumption for the bow shock is not valid for
$v_{250}\lesssim 2-3$. As a rough guide, one might still use the
results for a steady wind \citep{Wilkin96}, with the average spin down
luminosity over a period $t_{\rm bs}$. The exact shape of the bow
shock could, however, be somewhat different than that of a steady
wind.}  In our scenario, $2R_{\rm ext}$ (the radius for a one sided
outflow) is $\sim(1-2)R_{\rm bs}$, i.e. $R_{\rm bs}\sim(1-2)R_{\rm
ext}$ where $R_{\rm ext}\approx 6.4\times 10^{15}t_{\rm
col,5}d_{15}\;$cm, which is the case for our fiducial parameters.

Lower limits on the energy, $E\gtrsim 10^{44.5}\;$erg, and mass,
$M_{\rm ej}\gtrsim 10^{24.5}\;$g, of the outflow, and on the external
density, $n_{\rm ext}\gtrsim 10^{-2}\;{\rm cm^{-3}}$ have been derived
in \S \ref{emission} \citep[see also][]{Gelfand05}.  The values of
$E$, $M_{\rm ej}$ and $n_{\rm ext}$ scale linearly with each other,
$E\sim 10^{46.5}n_0\;$erg and $M_{\rm ej}\sim 10^{26.5}n_0\;$g. Note,
however, that the minimal allowed density, $n_0\sim 10^{-2}$, requires
$v_*\sim 2500\;{\rm km\;s^{-1}}$ and a similar kick velocity for
SGR~$1806-20$ at its birth. While this is an extremely high kick
velocity for typical radio pulsars, magnetars might have a
significantly higher kick velocity compared to ordinary neutron stars,
which might approach such high values \citep{DT92,TD95}. A lower kick
velocity would suggest that the true values of $E$, $M_{\rm ej}$ and
$n_{\rm ext}$ are larger than their lower limits by a factor of $\sim
100n_0\sim 100v_{250}^{-2}$.

An alternative mechanism for producing the external shell is the
ejection of a faster (and likely highly relativistic) component of the
ejecta which carries a small fraction, $f$, of its total energy, just
before (or in a slightly different direction relative to) the main
part of the ejecta which carries most of its energy and is only mildly
relativistic. The first component may be naturally identified with the
matter that is coupled to the radiation in the initial spike of the
GF, which is expected to reach a highly relativistic Lorentz factor
\citep{NPS05}. In order to obtain the ratio $R_{\rm dec}/R_{\rm
ext}\sim 4-5$ that is inferred from observations, this would require
$f\sim(R_{\rm dec}/R_{\rm ext})^{-3}\sim 10^{-2}$, i.e. that the
ultra-relativistic component would carry $\sim 1\%$ of the total
energy in the outflow.

\begin{figure}
\centerline{\includegraphics[width=9.0cm]{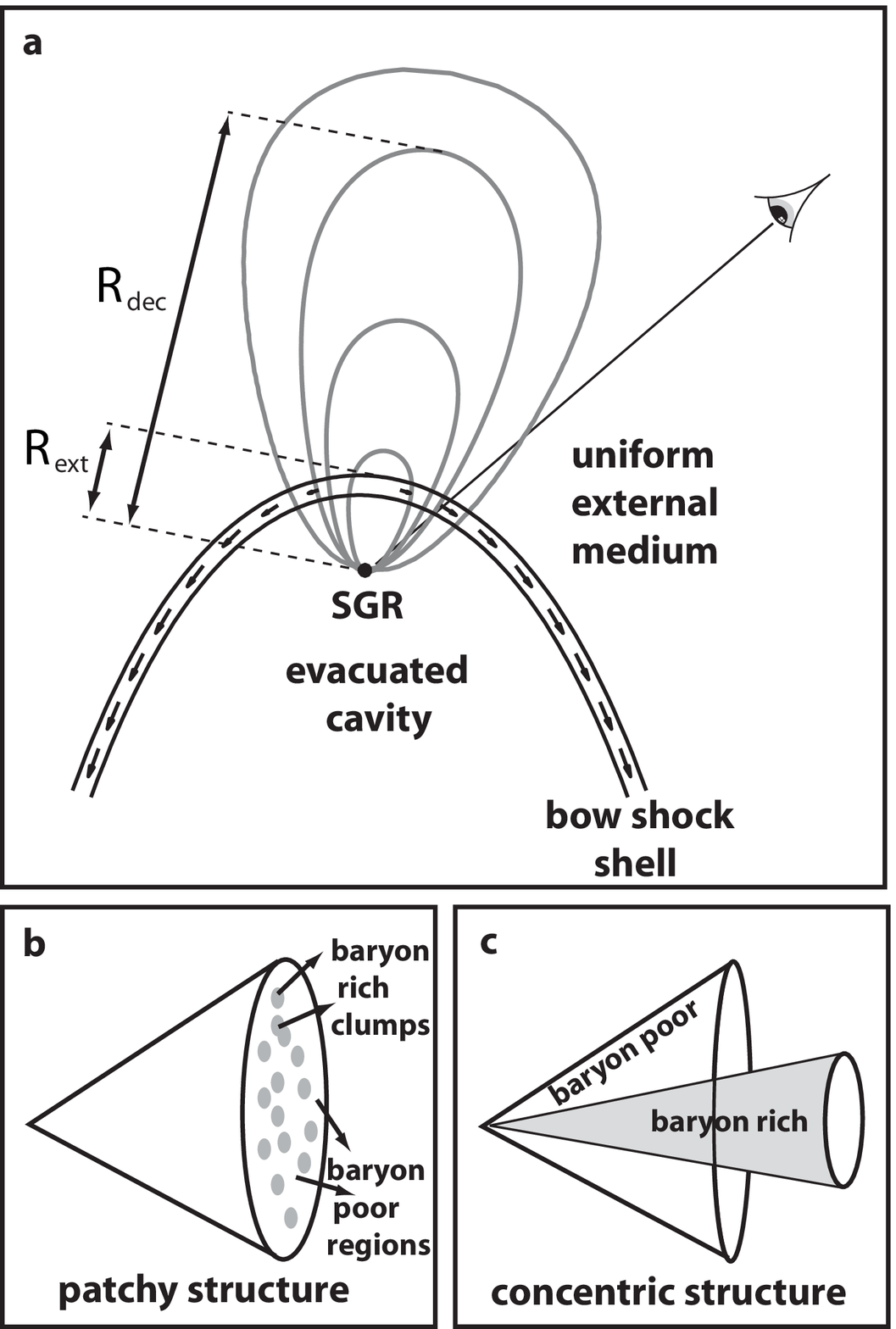}}
\caption{An illustration of the basic underlying geometry in our
  model. {\bf a)} A pre-existing shell surrounding a cavity (i.e. an
  evacuated region) is possibly formed due to the interaction of the
  SGR quiescent wind with the external medium, and the SGR's
  super-sonic motion relative to the external medium. The outflow from
  the SGR~$1806-20$ Dec. 27 giant flare was ejected mainly in one
  preferred direction, probably not aligned with the head of the bow
  shock (which is in the direction of the SGR's systemic motion). The
  ejecta collide with the external shell at a radius $R_{\rm ext}$ and
  then the merged shell of shocked ejecta and shocked swept up
  external shell keeps moving outward at a constant (mildly
  relativistic) velocity. As it coasts outward, it gradually sweeps
  the external medium until at a radius $R_{\rm dec}\sim(4-5)R_{\rm
  ext}$ it sweeps up a sufficient mass of external medium in order to
  decelerate significantly. At $R>R_{\rm dec}$ the structure of the
  flow gradually approaches the spherical self-similar Sedov-Taylor
  solution. {\bf b,c)} Most of the mass in the outflow was in baryons
  that were decoupled from most of the radiation, and our line of
  sight was baryon poor. This naturally occurs if there are separate
  baryon rich (radiation poor) and baryon poor (radiation rich)
  regions. Such regions might consist of small baryon rich clumps
  surrounded by baryon poor regions ({\it panel {\bf b}}), or might
  alternatively be part of a global large scale, possibly concentric,
  configuration ({\it panel {\bf b}}).}
\label{model}
\end{figure}

\begin{figure}
\centerline{\includegraphics[width=11.5cm]{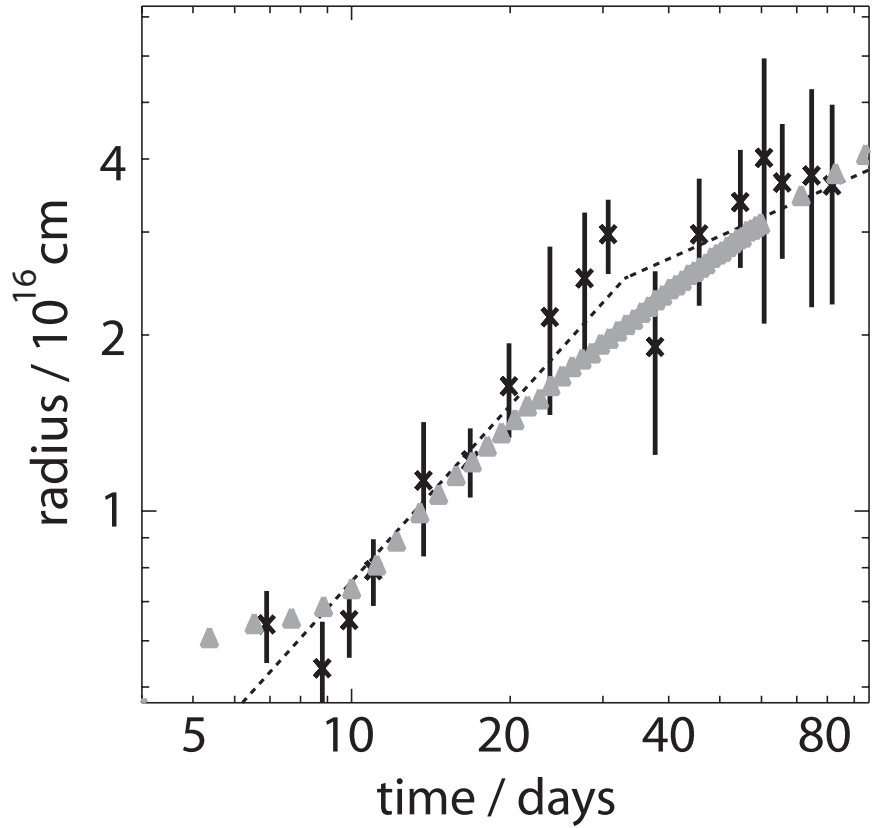}}
\caption{Temporal evolution of the observed size of the source from
  radio measurements ({\it asterices}) of \citet{Taylor05} assuming
  $d_{15}=1$, together with the source size from our numerical
  simulation ({\it triangles}) and for our simple analytic model ({\it
  dotted line}). Our numerical simulation featured the collision
  between the outflow ejected during the SGR giant flare and a
  pre-existing shell surrounding a cavity. The calculations were done
  in two-dimensional cylindrical coordinates for ten levels of
  refinement using the piecewise parabolic method (PPM) adaptive mesh
  refinement code FLASH.  The spherical initial configuration is as
  follows. In the inner region (outflow from the SGR, inner $5 \times
  10^{14}\;$cm) both a thermal energy of $E=10^{46}\;$erg and ejecta
  mass, $M_{\rm ej}$, are distributed uniformly; $M_{\rm ej}$ is
  selected so that $v=(2E/M_{\rm ej})^{1/2}\approx 0.4c$ (i.e. $M_{\rm
  ej}\approx 1.4\times 10^{26}\;$g).  The injected gas and surrounding
  ISM (with $\rho_{\rm ext} = 2 \times 10^{-24}\;{\rm g\;cm^{-3}}$)
  are characterized by a 5/3 adiabatic index. More details will be
  presented in Ramirez-Ruiz et al. (2005, in preparation).}
\label{hydro}
\end{figure}

The fact that most of the energy in the outflow was in mildly
relativistic ejecta, implies that the bulk of the ejecta was not
coupled to the radiation of the initial spike. This could occur if the
bulk of the outflow and of the radiation came out in different
directions (either in small local patches, or within some global,
possibly concentric, structure, as illustrated in panel {\bf b} and
{\bf c} of Fig. \ref{model}, respectively).  Thus, it is not even
obvious whether they occupied a comparable solid angle, which is
important when trying to compare their true energies.

Our line of sight must have been in a relatively baryon-poor (and
radiation rich) region, not only in order to see a bright initial
spike, but also since otherwise the high optical depth of the
electrons associated with the baryons in the outflow would have
obscured the first $\sim 30\;$s of the GF tail emission. This can be
seen as follows. If a shell of mildly relativistic proton-electron
plasma with velocity $v_0$ and isotropic equivalent mass $M_{\rm
ej,iso}$ is ejected during the initial spike, at $t\approx 0$, then
radiation emitted at time $t$ after the initial spike would reach this
shell at a radius $R(t)=v_0t/(1-v_0/c)$. The shell becomes optically
thin to the radiation from the tail of the GF the a time $t_*$ when
$\tau_T(t_*)=M_{\rm ej,iso}\sigma_T/[m_p 4\pi R^2(t_*)]=1$, i.e. at
$t_*=(1-v_0/c)v_0^{-1}(M_{\rm ej,iso}\sigma_T/4\pi m_p)^{1/2}$. This
gives $t_*\gtrsim 25\;$s from the lower limits we obtain for $M_{\rm
ej,iso}$ and for $\beta_{\rm min}\approx 0.7$ for a one-sided outflow
(a similar result is obtained for the spherical case). The presence of
$e^\pm$ pairs in the outflow or a large mass or velocity of the ejecta
would only increase $t_*$.

In order for the highly relativistic ejecta to form the external shell
into which the mildly relativistic ejecta collides, the two components
must have a significant overlap in solid angle by the time of the
collision, i.e. after the ultra-relativistic component decelerates to
Newtonian velocities. This could occur if the two components form
small patches, rather than a large scale coherent structure (such as a
baryon-rich core surrounded by a baryon-poor and radiation-rich outer
region). Alternatively, if the external shell was created by the bow
shock, then an energy ratio of $f\lesssim 10^{-2}$ is required in
order for the highly relativistic component not to alter the radio
emission considerably. This would be consistent with the very low
energy in the ultra-relativistic component that is expected for a pure
pair plasma \citep[or for a very low baryon loading;][]{NPS05}.

For a wide one-sided jet we find that the lower limit on the isotropic
equivalent kinetic energy in the outflow, $\gtrsim 5\times
10^{45}d_{15}^{2.5}\;$erg, is only a factor of
$\sim(3-4)d_{15}^{-0.5}$ smaller than the isotropic equivalent energy
radiated in the GF itself. This suggests that the two isotropic
equivalent energies are comparable. If the solids angles occupied by
the baryon-rich (radiation-poor) and by the radiation-rich
(baryon-poor) regions are comparable (which is not at all obvious),
then this would suggest that the true energies in the GF itself and in
the kinetic energy of the outflow are of the same order.

A much dimmer radio afterglow was detected following the 1998 Aug.  27
GF from SGR~$1900+14$ \citep{FKB99}, which despite the much sparser
data, shows similarities to the radio afterglow discussed here. This
suggests that our model might be applicable more generally. The spin
down luminosity $L$ of the two SGRs is comparable, and so is the time
at which the light curve started to decay steeply ($\sim 9-10\;$days),
suggesting that $R_{\rm bs}$ in each case is not very different.  This
would imply a similar $n_{\rm ext}v_*^2$. Under these assumptions, the
large difference in the radio luminosity (by a factor of $\sim 500$)
would be mainly a result of the much larger energy content carried by
the outflow of SGR~$1806-20$ immediately after the GF.

\acknowledgments This research was supported by US Department of
Energy under contract number DE-AC03-76SF00515 (J.G.) and by NASA
through a Chandra Postdoctoral Fellowship award PF3-40028 (E. R.-R.).
The software used in this work was in part developed by the
DOE-supported ASCI/Alliance Center for Astrophysical Thermonuclear
Flashes at the University of Chicago. Computations were performed on
the IAS Scheide computer cluster. B.M.G. and J.D.G. were supported by
NASA through LTSA grant NAG5-13032. The National Radio Astronomy
Observatory is a facility of the National Science Foundation operated
under cooperative agreement by Associated Universities, Inc.

\newpage

\newpage

\end{document}